\documentclass[conference]{IEEEtran}
\IEEEoverridecommandlockouts
\usepackage{cite}
\usepackage{algorithm,algorithmicx}
\usepackage{algpseudocode}
\usepackage[small,bf]{caption}
\usepackage[cmex10]{amsmath}
\usepackage{amssymb,latexsym,epsfig,subfigure,epic,amscd,mathrsfs,euscript,bm}
\usepackage{color}
\usepackage{booktabs}
\usepackage{array}
\usepackage{amsfonts}
\usepackage{amsopn}
\usepackage{graphicx}
\usepackage{url} 
\usepackage{empheq}
\usepackage{atbegshi}
\usepackage[inline]{enumitem}
\AtBeginDocument{\AtBeginShipoutNext{\AtBeginShipoutDiscard}}

\newtheorem{theorem}{\bf{Theorem}}
\newtheorem{remark}{\bf{Remark}}

\newtheorem{defn}{\bf{Definition}}

\newcommand{\bs}[1]{ \ensuremath{ \boldsymbol{#1} }}

\def\BibTeX{{\rm B\kern-.05em{\sc i\kern-.025em b}\kern-.08em
    T\kern-.1667em\lower.7ex\hbox{E}\kern-.125emX}}

\DeclareMathAlphabet\mathbfcal{OMS}{cmsy}{b}{n}

\begin{document}
%

\title{Fusion of Deep Neural Networks for Activity Recognition: A Regular Vine Copula Based Approach}

\author{\IEEEauthorblockN{Shan Zhang, Baocheng Geng and Pramod K. Varshney }
\IEEEauthorblockA{Department of Electrical Engineering and Computer Science \\
Syracuse University, Syracuse, NY 13244
\\
\{szhang60, bageng, varshney\}@syr.edu
}
\and
\IEEEauthorblockN{Muralidhar Rangaswamy}
\IEEEauthorblockA{Air Force Research Labs \\
Wright-Patterson Air Force Base, OH 45433
\\
Muralidhar.Rangaswamy@us.af.mil
}
}
\thanks{The work was supported by Air Force
Office of Scientific Research under Grant FA9550-17-1-0313 under the DDDAS
program. }
\maketitle


\maketitle

\begin{abstract}
In this paper, we propose regular vine copula based fusion of multiple deep neural network classifiers for the problem of multi-sensor based human activity recognition. We take the cross-modal dependence into account by employing regular vine copulas that are extremely flexible and powerful graphical models to characterize complex dependence among multiple modalities. Multiple deep neural networks are used to extract high-level features from multi-sensing modalities, with each deep neural network processing the data collected from  a single sensor. The extracted high-level features are then combined using a regular vine copula model. Numerical experiments are conducted to demonstrate the effectiveness of our approach.


\end{abstract}


\section{Introduction}
The recent development of small-size, low-power, multifunctional sensors and revolutionary advances in communication and computing technologies have resulted in many applications of sensor/information fusion. One such application is human activity recognition (HAR) that can be used for  health care, personal fitness, and border surveillance, etc. \cite{wu2011senscare, wu2013mobisens, iyengar2007detection}. 
The task of HAR is to detect and recognize human actions from the data provided by multiple sensors. HAR is naturally a classification task. Combining multiple sensing modalities can boost the classification performance. However, since each sensor carries a unique physical trait, sensor heterogeneity or incommensurability is the first critical challenge for multi-modal fusion.
Also, multiple sensor modalities tend to be dependent due to non-linear cross-modal interactions.  
In this paper, we study a classification problem by fusing multiple classifiers, one for each modality,
for the recognition of human actions. Thus, the data compatibility issues among multiple modalities can be avoided \cite{ngiam2011multimodal, lahat2015multimodal}. Moreover, the non-linear dependence is also addressed.

Most of the current multi-classifier fusion solutions for HAR rely on \textit{shallow} classifiers, such as Support Vector Machines (SVMs), Random Forests and Decision Trees, which employ handcrafted statistical  features extracted from each modality. 
%
%
The typical strategy for the fusion of these features is to combine the outputs 
obtained from multiple classifiers, where each classifier only takes the features of one modality \cite{zappi2007activity, stikic2008adl, li2010multimodal, banos2013human, guo2016wearable, nweke2019data}. 
Since the multivariate sensor observations are dependent, the outputs from the multiple classifiers are inherently dependent. This dependence can be non-linear. 
The design of an optimal fusion rule that can address this non-linear dependence becomes the key issue. In the aforementioned literature \cite{zappi2007activity, stikic2008adl, li2010multimodal, banos2013human, guo2016wearable, nweke2019data}, simple fusion rules, such as majority voting, naive Bayesian fusion, weighted average fusion, and Dempster-Shafer fusion, were used. 
None of these existing fusion rules in \cite{zappi2007activity, stikic2008adl, li2010multimodal, banos2013human, guo2016wearable, nweke2019data} exploit the non-linear dependence among multiple classifiers.
Moreover, designing and selecting robust features heavily relies on human experience and is time consuming. Also, only shallow features, such as mean, variance and amplitude, can be learned according to human expertise, which can be insufficient for more complex activities \cite{yang2015deep}.

Deep learning has had breakthroughs in a wide variety of unimodal applications, such as visual object recognition, natural language processing and object detection \cite{lecun2015deep}. Compared to the shallow classifiers, the deep classifiers can learn many more high-level features directly from raw data (or lightly processed data) and avoid the need for the design of handcrafted features. Recently, single deep learning models for HAR have been discussed in detail in the survey paper \cite{wang2018deep}. 

Very recently, multimodal deep learning methodologies for HAR have attracted some attention \cite{radu2018multimodal, moya2018convolutional}. 
Similar fusion strategies as discussed for multiple shallow classifiers can be applied to deep neural networks, based on the level where the fusion is performed: 
intermediate fusion with higher-level representations, referred to as high-level features, late fusion with decisions or probability scores. 
In \cite{radu2018multimodal, moya2018convolutional}, intermediate fusion strategies using Deep Neural Networks (DNNs) and Convolutional Neural Networks (CNNs) were studied, where a fully connected fusion layer was used to combine multiple DNNs or CNNs. As mentioned earlier, the data from multiple sensing modalities are non-linearly dependent. The fully connected fusion layer can learn this dependence in some manner. However, understanding and analyzing this non-linear dependence using the fully connected layer or another deep neural network is yet to be solved. 

Copula-based dependence modeling \cite{joe2014dependence} provides a flexible parametric characterization of the joint distribution of multiple sensor observations. It allows the separation of modeling different univariate marginals from modeling the multivariate (dependence) structure. The separated multivariate structure is referred to as a multivariate copula that can characterize non-linear or even more complex dependence.  It has been shown that copula-based fusion of multiple sensing  modalities can significantly improve the inference performance \cite{iyengar2007detection, he2012fusing, sundaresan2011location, iyengar2011biometric, he2015social}. However, for high-dimensional dependence structures, the underlying dependence can be more complex and the well defined multivariate copula functions, e.g., multivariate Gaussian copulas, multivariate Student-t copulas, may lack the ability to characterize the potential complex dependence  \cite{zhang2019fusion}. In this paper, we adopt a more general copula model, regular vine (R-Vine) copulas \cite{bedford2001probability, bedford2002vines, aas2009pair} that are constructed by decomposing a multivariate copula into a cascade of bivariate copulas. 

In this paper, we propose to leverage the DNNs and the R-Vine copula based dependence modeling for sensor-based recognition of human actions. More specifically, we use multiple DNNs to extract high-level features from multiple sensing modalities, where each DNN only takes the data from a single sensor. Different from the fusion strategy (using a fully connected fusion layer) in \cite{radu2018multimodal, moya2018convolutional}, we propose a probabilistic fusion methodology, R-Vine copula based fusion rule, that combines the extracted high-level features and characterizes the cross-modal dependence. Moreover, our proposed model is designed to improve the classification performance compared to the neural network based fusion method and adds interpretability in the sense that it explicitly explains the dependence structure of the extracted features from different modalities.

\section{Copula Theory}
\label{sec:CT}
A copula is a multivariate distribution with uniform marginal distributions. 
The unique correspondence between a multivariate copula and any multivariate distribution is stated in Sklar's theorem \cite{nelsen2013introduction} which is a fundamental theorem that forms the basis of copula theory. Also, for high-dimensional dependence structures, more flexible R-Vine copulas are constructed by decomposing a multivariate copula into a set of bivariate copulas, which will be discussed in Section \ref{sec:R-Vine_Fusion}.

\begin{theorem}[Sklar's Theorem]
The joint distribution function $F$ of random variables $x_1,\ldots,x_d$ can be cast as
\begin{equation}
\label{CopEq1}
F(x_1,x_2,\ldots,x_d) = C(F_1(x_1),F_2(x_2),\ldots,F_d(x_d)),
\end{equation}
where $F_1,\ldots,F_d$ are marginal distribution functions for $x_1,\ldots,x_d$.  If $F_m, m = 1, \ldots, d$ are continuous, $C$ is a unique $d$-dimensional copula. Conversely, given a copula $C$ and univariate Cumulative Distribution Functions (CDFs) $F_1,\ldots,F_d$, $F$ in \eqref{CopEq1} is a valid multivariate CDF with marginals $F_1,\ldots,F_d$.
\end{theorem}

For continuous distributions $F$ and $F_1,\ldots,F_d$, the joint Probability Density Function (PDF) of random variables $x_1,\ldots,x_d$ is obtained by differentiating both sides of \eqref{CopEq1}:
\begin{equation}
\label{CopEq2}
f(x_1,\ldots,x_d) \!=\! \Big(\!\!\prod_{m=1}^{d}f_m(x_m)\!\Big)c(F_1(x_1),\ldots,F_d(x_d)),
\end{equation}
where $f_1, \ldots, f_d$ are the marginal densities and $c$ is referred to as the density of the multivariate copula $C$ that is given by 
\begin{equation}
c(\mathbf{u}) = \frac{\partial^L(C(u_1,\ldots,u_d))}{\partial u_1,\ldots,\partial u_d},
\end{equation}
where $u_m=F_m(x_m)$ and $\mathbf{u} = [u_1,\dots,u_d]$. 
Note that $C(\cdot)$ is a valid CDF and $c(\cdot)$ is a valid PDF for uniformly distributed random variables $u_m$, $m = 1, 2, \ldots, d$. Since the random variable $u_m$ represents the CDF of $x_m$, the CDF of $u_m$ naturally follows a uniform distribution over $[ 0,1]$.


Various families of multivariate copula functions are presented in \cite{nelsen2013introduction}, such as elliptical and Archimedean copulas. Since different copula functions model different types of dependence, selection of copula functions to fit the given data is a key problem. 
Moreover, the \textit{dependence parameter} denoted by $\bs{\phi}$, contained in a copula function, is used to characterize the amount of dependence among $d$ random variables. Typically, $\bs{\phi}$ is unknown a \textit{priori} and needs to be estimated, e.g., using Maximum Likelihood Estimation (MLE) or Kendall's $\tau$ \cite{he2015heterogeneous}. Note that in general, $\bs{\phi}$ may be a scalar, a vector or a matrix. 

\section{Problem Statement}
\label{sec:PS}
\begin{figure}
\centering
\includegraphics[height=1.3in,width=!]{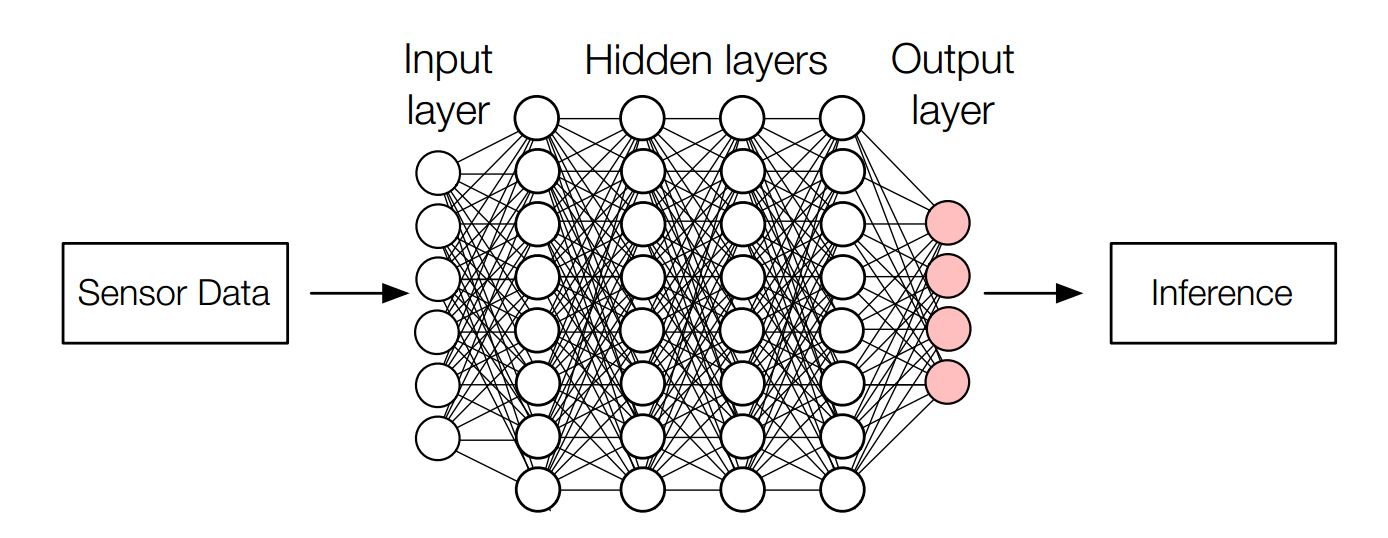}
\caption{A typical Deep Neural Network structure \cite{radu2018multimodal}.}
\label{fig:dnn}
\end{figure}

\begin{figure}[ht!]
\centering
\includegraphics[width=8cm]{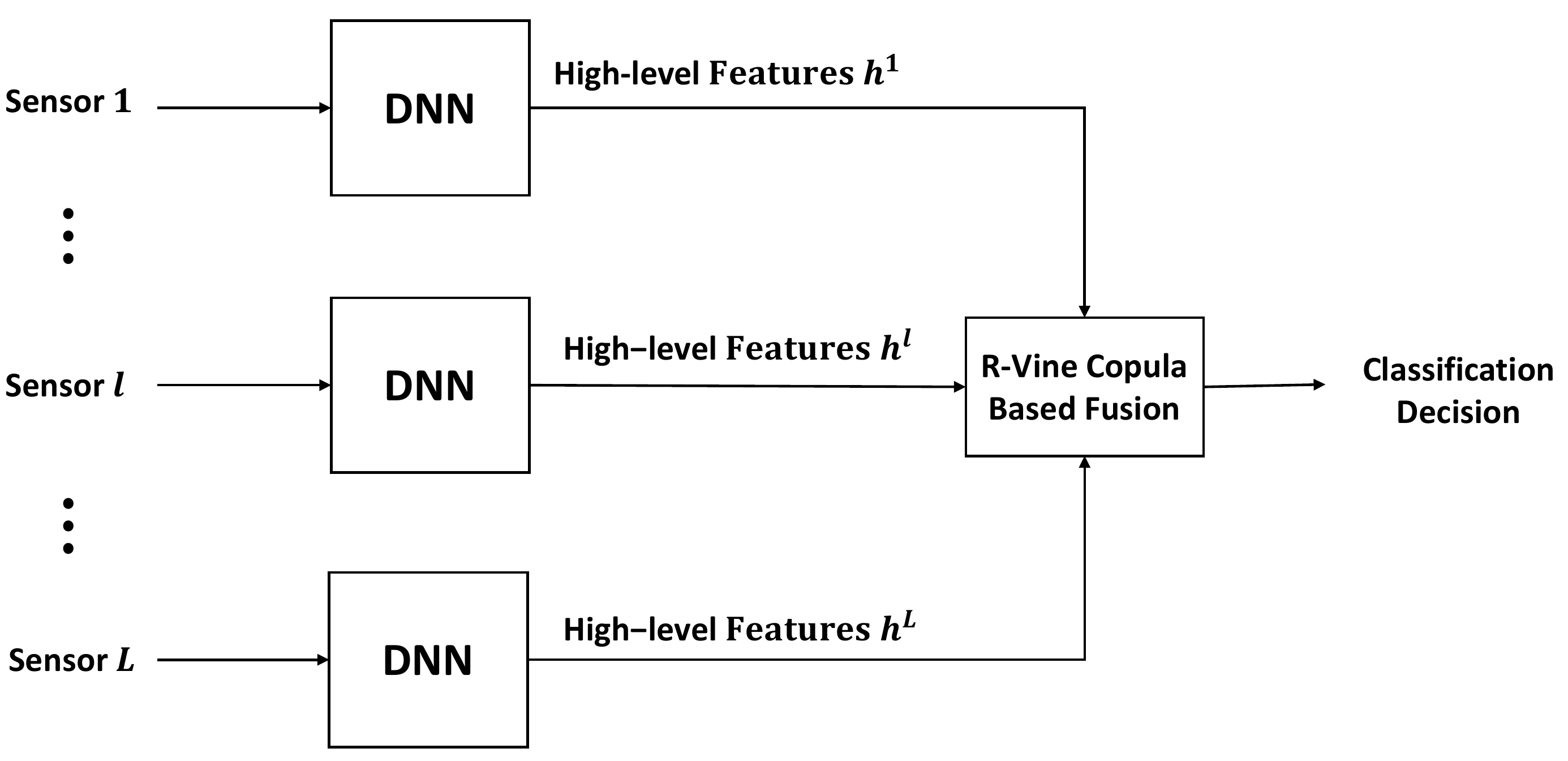} 
\caption{\footnotesize{R-Vine Copula Based Multi-modal DNN.}}
\label{fig:CF}
\end{figure}

Consider a supervised classification problem with $G$ classes. Let $\Omega = \{w_1, w_2, \ldots, w_G \}$ be the set of class labels. $L$ sensors make a set of observations regarding the object/event at time instant $n$, $\{ \mathbf x_{1n}, \mathbf x_{2n}, \ldots, \mathbf x_{Ln}, y_n\}$, where $n = 1, 2, \ldots$ and $\mathbf x_{ln} \in \mathbb R^{d_l^1 \times d_l^2}, d_l^1, d_l^2 \in \mathbb N, \mathbb N = [1, 2, \ldots]$ is the observation of sensor $l$ at time $n$. $y_n \in \Omega$ is the class label.
We assume that the sensor observations are continuous random variables that are conditionally independent and identically distributed (i.i.d.) over time. 
$L$ independent  pre-trained DNN classifiers are used to extract high-level features from each sensing modality. A typical DNN is shown in Fig. \ref{fig:dnn}. Compared to the traditional artificial neural networks, DNN is more capable of learning informative features from large amount of data. We use $\mathbf h_n^l \in \mathbb R^{1\times r_l}, r_l \in \mathbb N$ to represent the $n$th high-level feature vector extracted from sensor $l$. These high-level features are then combined using the R-Vine copula based fusion rule. We show the classification system to be studied in Fig. \ref{fig:CF}.

\begin{remark}
Note that for the sensor-based HAR, we use the DNNs to extract high-level features instead of CNNs. There are two main reasons. The first one is that compared to DNNs, CNNs are computationally more intensive. The second one is that the high-level features extracted from CNNs are generally high dimensional. Also, among these high dimensional features, a large number of features are irrelevant and redundant. Fusing all the features based on R-Vine copula models is computationally inefficient.
\end{remark}

Our aim is to determine the class label by combining the extracted high-level features. Assume that we have a training set with a total of $N$ feature vectors and the joint feature vector is
\begin{equation}
\label{eq:feature}
\mathbf h_n = [\mathbf h_{n}^1\,\, \mathbf h_{n}^2\,\, \ldots\,\, \mathbf h_{n}^L] \in \mathbb R^{1 \times (r_1+r_2+ \ldots + r_L)}, n = [N],
\end{equation}
where $[N]  = [1, 2, \ldots, N]$. In the following, for notational simplicity, we omit the superscripts of the feature vectors in \eqref{eq:feature} and let $\mathbf h_n = [h_{1n}, h_{2n}, \ldots, h_{Kn}], K = r_1+r_2+ \ldots + r_L$.

Using Bayes' theorem, the posterior probability of class $w_i$ given the joint high-level feature vectors is given as:
\begin{equation}
P(w_i | \mathbf h) = \frac{f(\mathbf h | w_i) P(w_i)}{f(\mathbf h)} \propto f(\mathbf h | w_i) P(w_i), 
\end{equation}
where $\mathbf h = [\mathbf h_1, \ldots, \mathbf h_N]$, $f(\mathbf h | w_i)$ is the joint likelihood function and $P(w_i)$ is the  prior probability for class $w_i$. If the class prior probabilities are not known, it is commonly assumed that the classes are equally likely. The class label $w_0$ is determined by choosing the label with highest posterior probability, which is given by
\begin{equation}
w_0 = \text{arg} \,\, \underset{w_i \in \Omega}{\text{max}} P(w_i | \mathbf h).
\end{equation}

Since $f(\mathbf h)$ is a constant for all the classes, the main problem is how to model and maximize $f(\mathbf h | w_i)$ under unknown multivariate dependence. In the following section, we will use R-Vine copulas to model the joint likelihood function.

\section{R-Vine Copula Based Fusion of Multiple Deep Neural Networks}
\label{sec:R-Vine_Fusion}
In this section, we present the R-Vine copula based fusion rule. Our goal now is to find the joint PDFs of feature vectors $\mathbf h$ under each class.
According to Sklar's theorem (Section \ref{sec:CT}), the joint PDF can be separated into its marginals and the dependence structure that is fully characterized by the copula (see \eqref{CopEq2}). Therefore, we have
\begin{equation}
f(\mathbf h | w_i) = \prod_{n = 1}^{N}\Big(\prod_{k=1}^{K}f_k(h_{kn} | w_i)\Big) c_i\Big(\mathbf F^{i}(\mathbf h_n) | \bs{\phi}_i \Big),
\end{equation}
where $f_k(h_{kn} | w_i), k = 1, \ldots, K$ are the marginal PDFs and $\mathbf F^{i}(\mathbf h_n) = [F^{i}_1(h_{1n}), \ldots, F^{i}_K(h_{Kn})]$ denotes all the marginal CDFs at time instant $n$ under class $w_i, w_i \in \Omega$. Moreover, 
$c_i$ is the copula density function for class $w_i$ and $\bs{\phi}_i$ is the corresponding parameter set.

Since we have no knowledge of the joint distributions of the extracted high-level features, the marginal PDFs, marginal CDFs, copula density functions and their corresponding parameters need to be estimated using the training dataset.  The estimation of the marginal distributions and optimal copula density functions for all the classes is similar. Therefore, the class index $i$ will be omitted for now to simplify notations. 

The marginal PDFs can be estimated non-parametrically using kernel density estimators \cite{wassermann2006all} that provide a smoothed estimate of true density by choosing the optimal bandwidth so that an accurate estimate is achieved. 
Further, the marginal CDFs can be determined by the Empirical Probability Integral Transforms (EPIT) \cite{he2012fusing}. The estimate of $F_k(\cdot)$ is given as
\begin{equation}
\hat{F}_k(\cdot) = \frac{1}{N} \sum_{n=1}^{N} \bs{1}_{h_{kn} < \{ \cdot \}},
\label{empCDF}
\end{equation}
where $\bs{1}_{\{\cdot\}}$ is the indicator function. 

Next, we discuss how to construct and find the optimal multivariate copula $c^*$ using R-Vine copula models, which was introduced by Bedford and Cooke in \cite{bedford2001probability, bedford2002vines}. 

\subsection{R-Vine copula Models}
Before we give the formal definition of the R-Vine copula, we first introduce R-Vine. A $d$ dimensional R-Vine $\mathcal{V} = (T_1, \ldots, T_{d-1})$ is a nest of $d-1$ trees, where the edges of the tree $T_i$ are the nodes of the tree $T_{i+1}$, and  if two edges in tree $T_i$ share a common node, they are connected in tree $T_{i + 1}$. By specifying bivariate copulas on each of the edges, we obtain R-Vine copula and it is defined as follows. 
\begin{defn}[R-Vine Copula]
\label{def:R-VineC}
An R-Vine copula consists of three parts, denoted by $(\mathbf{F}, \mathcal{V}, \mathbf{B})$, where
\begin{enumerate}
\item $\mathbf{F} = [F_1, F_2, \ldots, F_d]^T \in [0, 1]^d$ is a vector with uniform marginals.
\item $\mathcal{V}$ is a $d$-dimensional R-Vine.
\item $\mathbf{B} = \{ C_{\mathfrak{C}_{e, a}, \mathfrak{C}_{e, b} | D_e} \, | \, e = \{a, b \}  \in E_i, i = 1, 2, \ldots, d - 1; \bs \phi \}$ is a set of bivariate copulas with a set of parameters $\bs \phi$, where $E_i$ is the edge set for tree $T_i$, $\mathfrak{C}_{e, a}$, $\mathfrak{C}_{e, b}$ are the conditioned nodes of the edge $e$ and $D_e$ is the conditioning node set of the edge $e$.
\end{enumerate}
\end{defn}

Using the R-Vine copula model and Sklar's theorem, the joint PDF of a $K$-dimensional observation vector $\mathbf z = [z_1, \ldots, z_K]$  is given by
\begin{align}
\label{eq:densityvine-S}
&f(\mathbf z | \mathcal{V}, \mathbf B, \bs{\phi}) = \prod_{k=1}^K f(z_{k}) \prod_{i=1}^{K-1}\prod_{e\in E_i} \times  \\ \nonumber
&c_{\mathfrak{C}_{e, a}, \mathfrak{C}_{e, b} | D_e}(F_{\mathfrak{C}_{e, a} | D_e}(z_{\mathfrak{C}_{e, a}} | \mathbf{z}_{D_e}),F_{\mathfrak{C}_{e, b} | D_e}(z_{\mathfrak{C}_{e, b}} | \mathbf{z}_{D_e}); \bs{\phi}), \nonumber 
\end{align}
where $e=\{a, b\}$ and $\mathbf{z}_{D_e}=\{z_{j} | j \in D_e\}$, $f(z_{k})$ is the marginal PDF of the observation of variable $k$, $k = 1, \ldots, K$. Moreover, the conditional distribution $F_{\mathfrak{C}_{e, a} | D_e}(z_{\mathfrak{C}_{e, a}} | \mathbf{z}_{D_e})$ can be obtained recursively tree by tree using the following equation \cite{joe1996families}:
\begin{equation}
\label{eq:FF}
\frac{\partial C_{\mathfrak{C}_{a, a_1}\!, \mathfrak{C}_{a, a_2}\! |D_a}\!\!\left(\!F_{\mathfrak{C}_{a, a_1}\! | D_a}\!(\!z_{\mathfrak{C}_{a, a_1}} \!| \mathbf{z}_{D_a}\!)\!, F_{\mathfrak{C}_{a, a_2}\! | D_a}\!(\!z_{\mathfrak{C}_{a, a_2}} | \mathbf{z}_{D_a}\!)\!\right)}{\partial F_{\mathfrak{C}_{a, a_2}\! | D_a}\!(\!z_{\mathfrak{C}_{a, a_2}}\! | \mathbf{z}_{D_a}\!)},
\end{equation}
where $e = \{a, b\} \in E_i$, $a = \{a_1, a_2\}$ and $b = \{b_1, b_2\}$ are the edges that connect $\mathfrak{C}_{e, a}$ and $\mathfrak{C}_{e, b}$ given the conditioning variables $D_e$. Similarly, we can obtain $F_{\mathfrak{C}_{e, b} | D_e}(z_{\mathfrak{C}_{e, b}} | \mathbf{z}_{D_e})$.

An illustrative 5-dimensional R-Vine copula example is shown in Fig. \ref{fig: example_RVine}. For variables $z_1, z_2, z_3, z_4, z_5$ and if they share the same dependence structure as in Fig. \ref{fig: example_RVine}, according to \eqref{eq:densityvine-S}, their joint PDF can be expressed as
\begin{equation*}
\begin{aligned}
&f(z_1, \!z_2, \!z_3, \!z_4, \!z_5)= \left [ \prod_{l=1}^5\! f(z_{l}) \right]\!\! \cdot \!c_{1,2}\! \cdot \!c_{2,3}\!\cdot \!c_{2,4}\! \cdot \!c_{3,5}\! \cdot \!c_{1,3 | 2} \\
&\cdot\! c_{3,4 | 2}\! \cdot \!c_{2,5 | 3} \! \cdot \!c_{1,4 | 23}\! \cdot \!c_{1,5 | 23}\! \cdot \!c_{4,5 | 123},
\end{aligned}
\end{equation*}
where the inputs for the bivariate copulas are omitted.

\subsection{Estimation of Optimal R-Vine copula}
The estimation of optimal R-Vine copula model for the joint feature vector $\mathbf h$ requires the selection of the R-Vine tree structure $\mathcal{V}$, the choice of copula families for the bivariate copula set $\mathbf B$ and the estimation of their corresponding parameters $\bs{\phi}$. 
To select the optimal R-Vine tree structure, we adopt the sequential maximum spanning tree algorithm in \cite{dissmann2013selecting}. This sequential method is based on Kendall's $\tau$. The sequential method starts with the selection of the first tree $T_1$ and continues tree by tree up to the last tree $T_{K-1}$. The trees are selected in a way that the chosen bivariate copula models the strongest pair-wise dependencies present which are characterized by Kendall's $\tau$. 
After the optimal R-Vine tree structure is selected, we need to define a bivariate copula family and estimate the optimal bivariate copulas that best characterizes the pair-wise dependencies.

Consider a library of copula, $\mathcal C = \{ c_m: m = 1, 2, \ldots, M\}$ Before estimating the optimal bivariate copula, the copula parameter set $\bs{\phi}$ is first obtained using MLE, which is given by
\begin{equation}
\label{eq:phi}
\hat{\bs{\phi}} \!=\! \text{arg}\max_{ \bs{\phi} } \sum_{n=1}^N{ \log c(\hat{F}_{k_1}(h_{k_1n}), \hat{F}_{k_2}(h_{k_2n})|\bs{\phi}) },
\end{equation}
where $(k_1, k_2), k_1, k_2 \in [1, 2, \ldots, K]$ is a connected pair in the selected R-Vine tree $\mathcal{V}$ and for simplification of notation, we omit the conditioning elements for conditional marginal CDFs. Note that the conditional marginal CDFs can be obtained recursively using \eqref{eq:FF}.

The best copula $c^*$ is selected from the copula library $\mathcal{C}$ using the Akaike Information Criterion (AIC) \cite{akaike1973information} as the criterion, which is given as
\begin{equation}
\text{AIC}_m =  - \sum_{n = 1}^{N} \log c_m(\hat{F}_{k_1}(h_{k_1n}), \hat{F}_{k_2}(\!h_{k_2n}) | \hat{\bs{\phi}}_m) + 2q(K),
\end{equation}
where $q(K)$ is the number of parameters in the $m$th copula model. Also, the conditioning elements for conditional marginal CDFs are omitted.

The best copula $c^*$ is
\begin{equation}
\label{eq:bestc}
c^* = \arg \min_{c_m \in \mathcal{C}} \text{AIC}_m.
\end{equation}

\section{Numerical Results}
In this section, we demonstrate the efficacy of our proposed R-Vine copula based methodology for the fusion of multiple DNNs. To show the superiority of our proposed fusion scheme, we also compare our result with the classification performance obtained by using the following schemes:

\begin{itemize}
    \item Single modality without fusion: Feed the raw data into a DNN classifier.
    \item Data-level fusion: Concatenate all the raw data from different modalities into one input vector and feed it into a DNN classifier.
    \item Fully connected layer fusion: Concatenate the extracted features into one feature vector and use a fully connected fusion layer to achieve a final classification decision.
\end{itemize}


\begin{figure}
\centering
\includegraphics[height=1.4in,width=!]{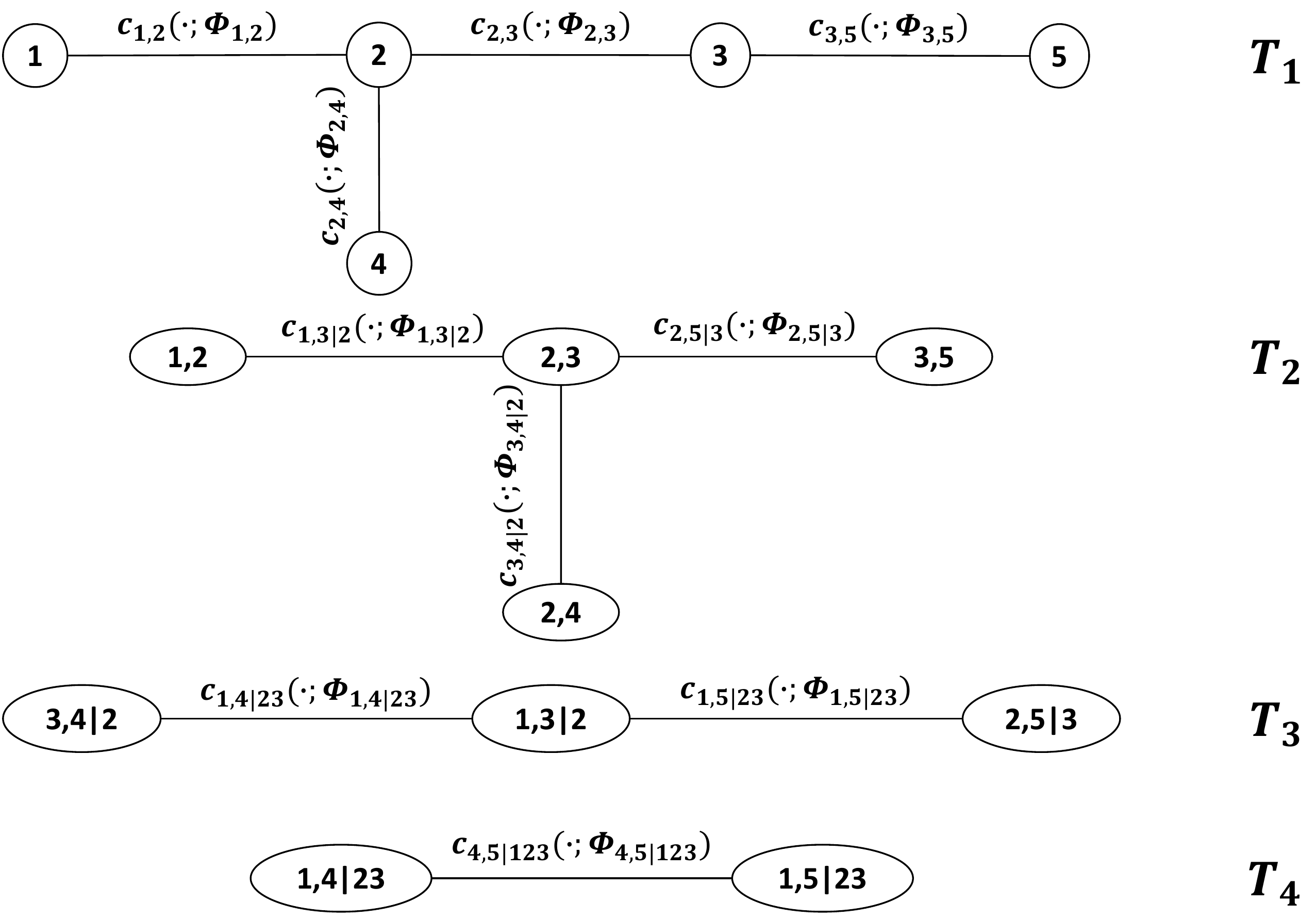}
\caption{An example R-Vine copula for five variables.}
\label{fig: example_RVine}
\end{figure}

\begin{table}[thb]
\centering
\caption{\textbf{STISEN}: $F_1$ scores for Watch-DNN, Phone-DNN, Fully-connected layer fusion, Data-level fusion, R-Vine copula fusion.}
\label{table:1}
\begin{tabular}{|c|c|}
\hline
   Model                   &  $F_1$ score        \\ 
\hline
Watch-DNN           & 71.4\%           \\ 
\hline
Phone-DNN            & 70.2\%        \\     
\hline        
Fully-connected layer fusion             & 78.0\%        \\ 
\hline
Data-level fusion          & 79.3\%          \\ 
\hline
R-Vine copula fusion           &88.6\%           \\ 
\hline
\end{tabular}
\end{table}

\begin{table}[thb]
\centering
\vspace{0.2cm}
\caption{\textbf{ANGUITA}: $F_1$ scores for Accelerometer-DNN, Gyroscope-DNN, Fully-connected layer fusion, Data-level fusion, R-Vine copula fusion.}
\label{table:4}
\begin{tabular}{|c|c|}
\hline
   Model                   &  $F_1$ score         \\ 
\hline
Accelerometer-DNN           & 87.8\%             \\ 
\hline
Gyroscope-DNN            & 72.9\%        \\     
\hline        
Fully-connected layer fusion             & 91.9\%        \\ 
\hline
Data-level fusion          & 88.3\%          \\ 
\hline
R-Vine copula fusion           & 92.8\%          \\ 
\hline
\end{tabular}
\end{table}

\begin{table*}[ht]
\centering
\vspace{0.4cm}
\caption{\textbf{STISEN}: Confusion matrix for R-Vine copula based fusion.}
\label{table:2}
\begin{tabular}{|c|c|c|c|c|c|c|c|}
\hline
 & Sit  & Stand &Walk & Stairsup & Stairsdown  & Bike   & Recall              \\ \hline
Sit           & 498 & 0  &0  & 0 &2 &0  & 99.6\%            \\ \hline
Stand            & 0 & 454  &0 & 0 &46 &0 & 90.8\% \\ \hline
Walk             & 0 & 0  &402 &43 &17 &38 &80.4\%       \\ \hline
Stairsup             & 0 &0  &24 & 408 &50 &0  & 81.6\%         \\ \hline
Stairsdown            & 0 & 0 &39 &53 &408 &0  & 81.6\%         \\ \hline
Bike            & 0 & 0  &8 &4 &2 &486  & 97.2\%         \\ \hline
Precision &100.0\%   &100.0\%  &85.0\%  &80.3\%  &77.7\%  &89.7\% & 88.6\%   \\ \hline
\end{tabular}
\end{table*}

\begin{table*}[ht]
\centering
\caption{\textbf{ANGUITA}: Confusion matrix for R-Vine copula based fusion.}
\label{table:3}
\begin{tabular}{|c|c|c|c|c|c|c|c|c|}
\hline
                      & W  & WU &WD & Si & St  & L  & Recall            \\ 
                      \hline
Walking           & 276 & 0  &17  & 3 &0 &0 &93.2\%             \\ 
\hline
Walking-upstairs            & 6 & 259  &0 & 3 &3 &0 & 95.6\%.        \\     
\hline        
Waking-downstairs             & 8 &0  &211 & 0 &1 &0 & 95.9\%        \\ 
\hline
Sitting           & 3 & 6 &1 &248 &33 &0 & 85.2\%           \\ 
\hline
Standing          & 3 & 3  &2 &26 &298 &0 &87.4\%          \\ 
\hline
Laying & 2 & 0  &0 &0 &6 &329 & 97.63\%        \\ 
\hline
Precision & 92.6\% & 96.6\%  &91.3\% &88.6\% &87.4\% &100.0\% &92.8\%        \\ 
\hline
\end{tabular}
\end{table*}

\subsection{Datasets}
We select two publicly available datasets that contain multi-modality sensor readings for the recognition of human activities. 

\textbf{STISEN} Heterogeneity Activity Recognition Dataset, collected by Stisen et al. \cite{stisen2015smart}, contains the sensor readings from two modalities: smartphone and smart watch. Each modality is equipped with two motion sensors, accelerometer and gyroscope. There are $6$ classes (`Sit', `Stand', `Walk', `Stairsup', `Stairsdown', `Bike') to be classified. We focus on the fusion of phone and watch modalities. Each of the two motion sensors produces a three-dimensional data vector, making each data sample contain 6 attributes in total. We select the data captured by Samsung Galaxy S3 mini phone and Samsung Galaxy Gear watch, where the data samples were sampled at a rate of $100$ Hz. $9000$ samples from each modality are used to train and test DNN models for feature selection, and another $9000$ samples are used to train and test the R-Vine copula based fusion methodology.

\textbf{ANGUITA} Human Activity Recognition Using Smartphone Dataset, collected by Anguita et al. \cite{anguita2013public}, contains accelerometer and gyroscope three-dimensional sensor data. It was collected from 30 volunteers who performed six different activities (`Walking', `Walking-upstairs', `Walking-downstairs', `Siting', `Standing', `Laying'). We focus on the fusion of accelerometer and gyroscope modalities. These sensor data were sampled at a rate of $50$ Hz, and were separated into windows of $128$ values. Each window has $50$\% overlap with the previous window. The $128$-real value vector in each window stands for one sample for each activity. 

\subsection{Classification Accuracy}
We use $F_1$ score as the classification performance metric, which is given by
\begin{equation}
F_1 = \frac{2}{|\Omega|}\, \underset{w}{\sum}\, \frac{\text{precision}_w \times \text{recall}_w}{\text{precision}_w + \text{recall}_w},
\end{equation}
where $\text{precision} = \frac{TP}{TP + FP}$ and $\text{recall} = \frac{TP}{TP+FN}$. Here, TP, FP and FN denote true positive, false positive and false negative, respectively. $F_1$ score is robust to unbalanced distributions of data samples across classes.

Table \ref{table:1} and Table \ref{table:4} show the $F_1$ scores comparing the five classification schemes: two single modalities based DNN classifiers and three multi-modal fusion based DNN classifiers for the \textbf{STISEN} and \textbf{ANGUITA}  datasets, respectively. As we can see, fusion based schemes perform better than single modality based schemes. Also, our proposed R-Vine copula based fusion methodology performs better than using the data-level fusion scheme and fully connected fusion layer scheme. Our proposed methodology achieves an overall $88.6\%$ and $92.8\%$ $F_1$ scores for the \textbf{STISEN} and \textbf{ANGUITA}  datasets, respectively. Moreover, we can see that for phone and watch modalities, the R-Vine copula based fusion scheme achieves higher performance improvement compared to accelerometer and gyroscope modalities. This is because of the fact that the accelerometer and gyroscope are less dependent while the phone and watch are highly dependent.

It should be noted that the training of R-Vine copula models requires less number of training samples compared to the training of a fully connected fusion layer or another DNN used for fusion. In Table \ref{table:1} and Table \ref{table:4}, the performance of our R-Vine copula scheme is obtained by using a total of $N = 1200$ feature samples. However, the fully connected fusion layer based scheme requires a total of $N = 6000$ feature samples. 

\begin{figure}[ht!]
\centering
\includegraphics[width=6.8cm]{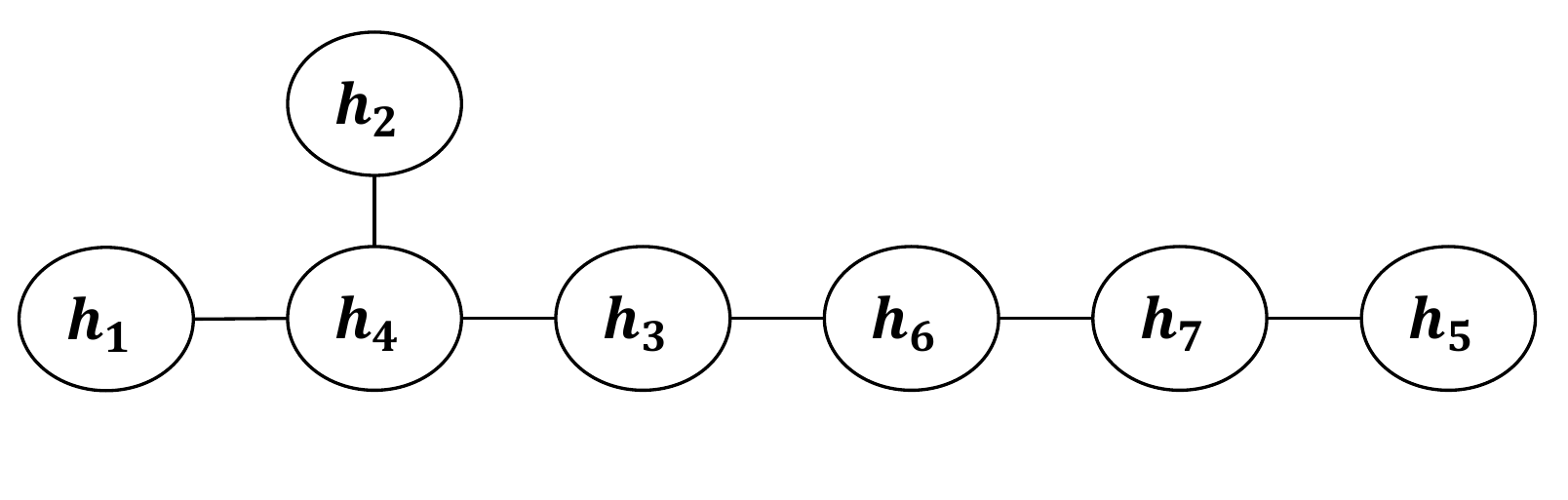} 
\caption{\footnotesize{First level dependence structure for activity `Walking-upstairs'.}}
\label{fig:dp}
\end{figure}

In Fig. \ref{fig:dp}, we show the first level dependence structure (first tree of the R-Vine copula model; see Fig. \ref{fig: example_RVine} as an example) of the extracted features using our proposed R-Vine copula based fusion method for activity `Walking-upstairs' in \textbf{ANGUITA} dataset. Here, features $\mathbf h_1, \mathbf h_2, \mathbf h_3, \mathbf h_4$ are from the accelerometer while the features $\mathbf h_5, \mathbf h_6, \mathbf h_7$ are from the gyroscope. As we can see that, features $\mathbf h_3$ and $\mathbf h_6$ are highly correlated and the two modalities accelerometer and gyroscope are correlated mainly via these two features. Using the knowledge of intra-modal and cross-modal feature dependencies, we can trace back and find where these dependent features originated from, which would yield the reduction of training data needed in DNNs. Furthermore, we are able to understand the correlation among the raw data from different modalities. The R-Vine copula based fusion method adds interpretability of the model which explicitly provides the dependence structure for features from different modalities, compared to the neural network based fusion which is totally a `black-box' model.

Table \ref{table:2} and Table \ref{table:3} show the confusion matrices using the R-Vine copula based fusion scheme for the \textbf{STISEN} and  \textbf{ANGUITA} datasets, respectively. As we observe, the fusion of phone and watch modalities achieves perfect classification for static activities (`Sit' and `Stand'). Also, the fusion of the accelerometer and gyroscope from the smartphone achieves significantly accurate classification for moving activities (e.g., `Walking', `Laying').

\section{Conclusion}

In this paper, an R-Vine copula based feature fusion approach was presented to perform activity recognition using multi-modal sensor observations. The features of each modality were extracted via a DNN and afterwards, an R-Vine copula model was constructed to capture the dependencies of intra-modal and cross-modal features. The procedures of model construction involve selecting the optimal R-Vine tree structure, obtaining the copula parameter set $\bs \phi$, and choosing the best copula $c^*$. Experiments on two human activity datasets demonstrated the efficiency of our proposed method compared to neural network based data/feature fusion, in terms of high prediction accuracy, less number of training samples required and dependence interpretability. In the future, we aim to address the problem of feature selection while constructing copula fusion model to achieve the balance of computation efforts and classification performance.



\bibliographystyle{IEEE}
\bibliography{reff_DNN,ref_HAR,refcopulaJ}

\end{document}